# A Novel Procedure For The Assessment of The Seismic Performance Of Frame Structures by means of Limit Analysis

A. Greco[a], F. Cannizzaro[a], A. Pluchino[b]

[a]Department of Civil Engineering and Architecture,
University of Catania, viale A. Doria 6, Catania, Italy
[b]Department of Physics and Astronomy "E.Majorana" University of Catania
and Sezione INFN of Catania, via S.Sofia 64, Catania, Italy
email: annalisa.greco@unict.it,
francesco.cannizzaro@dica.unict.it, alessandro@pluchino.it,

## Abstract

Limit analysis is a computationally efficient tool to assess the resistance and the failure mode of structures but does not provide any information on the displacement capacity, which is one of the concepts which most affects the seismic safety. Therefore, since many researchers did not consider limit analysis as a possible tool for the seismic assessment of structures, its widespread employment has been prevented. In this paper this common belief is questioned and the authors show that limit analysis can be useful in the evaluation of the seismic performance of frame structures. In particular, to overcome the limitation on the possibility to evaluate the displacements of a structure based on a limit analysis approach, an approximated capacity curve is reconstructed. The latter is based on a limit analysis strategy, which takes into account the second order effects, and evaluates the displacement capacity considering a post-peak softening branch and a threshold on the allowed plastic rotations. Then, based on this simplified capacity curve, an equivalent single degree of freedom system is defined in order to assess the seismic performance of frame structures. The proposed simplified strategy is implemented in a dedicated software and the obtained results are validated with well-established approaches based on nonlinear static analyses, showing the reliability and the computational efficiency of this methodology

**Corresponding author**: A.Greco, e-mail: annalisa.greco@unict.it, tel: +390957382251

# 1 Introduction

Modern approaches to assess the safety of buildings with respect to seismic excitations are related to three main concepts, namely resistance, ductility capacity and failure mode. In particular, the seismic performance of a building is considered acceptable if it guarantees a considerable resistance to lateral loading and it is capable of withstanding significant horizontal displacements. Furthermore the expected failure mechanism should dissipate as much energy as possible during the seismic event, i.e. it has to be a global failure mode. The numerical tools adopted to analyse a structure should reflect the possibility to adequately evaluate all these features.

Although nonlinear dynamic analysis is the best tool to simulate the seismic response of a structure, two main drawbacks have been limiting its wide employment in practical engineering. In particular, the uncertainty in the definition of appropriate ground motion inputs and the computational burden associated to nonlinear time histories, delayed the introduction of this analysis strategy as a reliable tool to assess the seismic performance of a building. Recently, due to the advances in the computing machines, its employment in practical engineering developed, with particular reference to structures with concentrated nonlinearities (e.g. seismically isolated buildings).

An alternative approach to the seismic assessment of buildings is based on the employment of nonlinear static analysis. Such a strategy, although not providing any insight on the cyclic behaviour of a structure, is able to provide information on the resistance, on the ductility and on the failure mechanism of a structure, through the construction of a capacity curve. Usually this curve is post-elaborated, determining a Single Degree of Freedom (SDOF) system (usually with elastic-plastic behaviour), which is considered equivalent to the building, on which verifications are made considering an inelastic spectrum [1]. This approach is the most widely adopted strategy to assess the seismic performance of a structure and it is acknowledged as an objective procedure for the evaluation of the seismic safety of a building and therefore adopted by the main structural codes [2,3]. The achievement of limit states is usually associated to particular coded conditions, which conventionally imply that a structure reveals a particular damage state. Usually, the limit states are associated to specific values of the base shear (peak or residual values) or, for frame structures, to the achievement of yielding and ultimate rotations in one of the plastic hinges. Nevertheless, such a strategy usually requires to perform the analyses considering different directions and force distributions, which increase significantly the amount of simulations to be run and, as a consequence, the computational cost associated to this methodology. Although nonlinear static analyses can be combined with sophisticated model in terms of constitutive laws and discretization, when it comes to complex structures, it is common to simplify the numerical models, for example limiting to uniaxial elements, or adopting elastic-perfectly plastic hinges, to facilitate the management of the numerical models. However, nonlinear static analyses represent at the moment the standard strategy for performance based seismic assessment of structures.

Limit analysis, which was introduced as a promising approach to evaluate the collapse conditions of a structure in terms of resistance and failure mechanism, did not find extensive applications for the assessment of the seismic performance of a building. This was mainly due to the intrinsic limitations of the method which can identify the resistance to lateral loadings and the collapse mechanism, but is not able to evaluate the magnitude of the displacements associated to the achievement of the ultimate conditions. In addition, the assumption of rigid perfectly plastic behaviour prevents the possibility to grasp the presence of softening in a structure as the magnitude of displacement and damage increases.

In the last decades the advances on limit analysis were mainly employed for academic studies, rather than finding a spread end use for practical purposes. An exception is the assessment of the resistance and the collapse mechanism of masonry walls subjected to out-of-plane lateral loadings, for which only recently reliable structural models have been proposed [4].

On the other hand, with reference to the evaluation of the ultimate conditions of structures, copious proposals employing limit analyses were made in the last decades [5-15]. Within this framework, and in particular with reference to seismic load conditions, the authors recently proposed to couple the method of the combination of elementary mechanisms with evolutionary algorithms [16,17] both for regular and irregular frames [18].

The main obstacles for limit analysis to be employed as a reliable numerical tool for the assessment of the seismic safety of structures are its unsuitableness in grasping the base shear reduction as the magnitude of displacement increases, as well as the impossibility of quantifying the plastic deformations at the parts of the structures involved in the failure mode. So, a question arises: can limit analysis be employed as a valid numerical strategy to assess the seismic performance of a structure?

In this paper, taking advantage of previous achievements and with the aim of providing an original contribution, the authors try to positively answer the question. The first of the mentioned drawbacks, that is relating the magnitude of displacements with the corresponding resistance of a structure considering a limit analysis approach, was developed in [19,20] considering the second order effects. The quantification of displacements using a limit analysis approach was performed with reference to geotechnical problems [21], but, to the best authors' knowledge, it was never proposed for civil buildings and in particular framed structures.

In the present study, the authors identify a simplified but reliable bilinear capacity curve for frame structures with lumped plasticity, introducing two main contributions in the field. The first one consists in the evaluation of the second order effects for the identification of a post peak softening branch; this result, although not novel, is here inferred for the first time in combination with the limit analysis methodology proposed by Neal and Symonds [22,23]. The second contribution of this study consists in the identification of the limit states on the post-peak branch of the simplified capacity curve, by means of an original computation of the plastic deformations at the hinges here proposed. Once the bilinear capacity curve is conveniently truncated, classic approaches to check the seismic performance of the structure are applied. The strategy is implemented in dedicated software developed within the agent-based programming language Netlogo [24]. The proposed methodology, which is very simple and decisively parsimonious from a computational point of view, with respect to the nonlinear static approach, is verified with pushover analyses. The reported results, although not exhaustive, open new perspectives on the possibility to wider scope employment of limit analysis in earthquake engineering.

## 2 Limit analysis of frames accounting for second order effects

Limit analysis of frame structures can be performed according to different strategies. Within the framework of the kinematic limit analysis theorem, the combination of elementary mechanisms approach [22,23] is here used. The authors recently employed this method, in combination with evolutionary algorithms, to assess the lateral resistance of regular [16,17] and irregular [18] frame structures. In the following subsections this strategy is briefly recalled at first; then an extension of the methodology to account for the second order effects is proposed.

*2.1 The method of combination of elementary mechanisms for frame structures*

In the present study planar regular frames with columns clamped at the base are considered. These are characterized by the number of floors $N_f$ and the number of columns $N_c$.

By considering the *i*-th floor and the *j*-th column of the frame, the plastic moments of the structural members are assumed to be $M_{b,ij}$ for beams and $M_{c,ij}$ for columns. Numbering of the nodes has been introduced as well ($N_{ij}$).

The frame may be loaded, at each floor, by concentrated horizontal forces $F_i$ and vertical distributed loads $q_{ij}$. Plastic hinges in the collapse mechanism can therefore be located in *s* "critical sections" correspondent to each joint and to a certain section of each beam, which can vary along the span. It is worth to point out that in the case of more than two members converging in a joint, a different critical section must be considered for each member (Figure 1).

The present paper deals with a kinematic plastic analysis and, following the approach originally proposed by Neal and Symonds [22,23] and already developed by the authors in previous works [16-18], takes into account global collapse mechanisms obtained by means of linear combination of three elementary ones: floor, beam and node mechanisms.

In seismic analysis of structures, the horizontal forces are considered variable while the vertical loads are assumed to be distributed and of constant value, therefore for each collapse mechanism, obtained combining the elementary ones, the virtual work theorem states:

$$\lambda_o W_{extH} + W_{extV} = W_{int} \quad (1)$$

where $W_{extH}$ and $W_{extV}$ represent the work done by the horizontal forces and the vertical permanent load, respectively, and $W_{int}$ is the internal work concentrated in the plastic hinges. The value of the load multiplier $\lambda_o$ is therefore given by:

$$\lambda_o = \frac{W_{int} - W_{extV}}{W_{extH}} \quad (2)$$

It is worth to mention that the location of possible along axis plastic hinges in the beams is computed a priori according to [19].

As it is very well known, analysing all the possible combinations of elementary mechanisms, the minimum value of $\lambda_o$ must be sought in order to obtain the real collapse load.

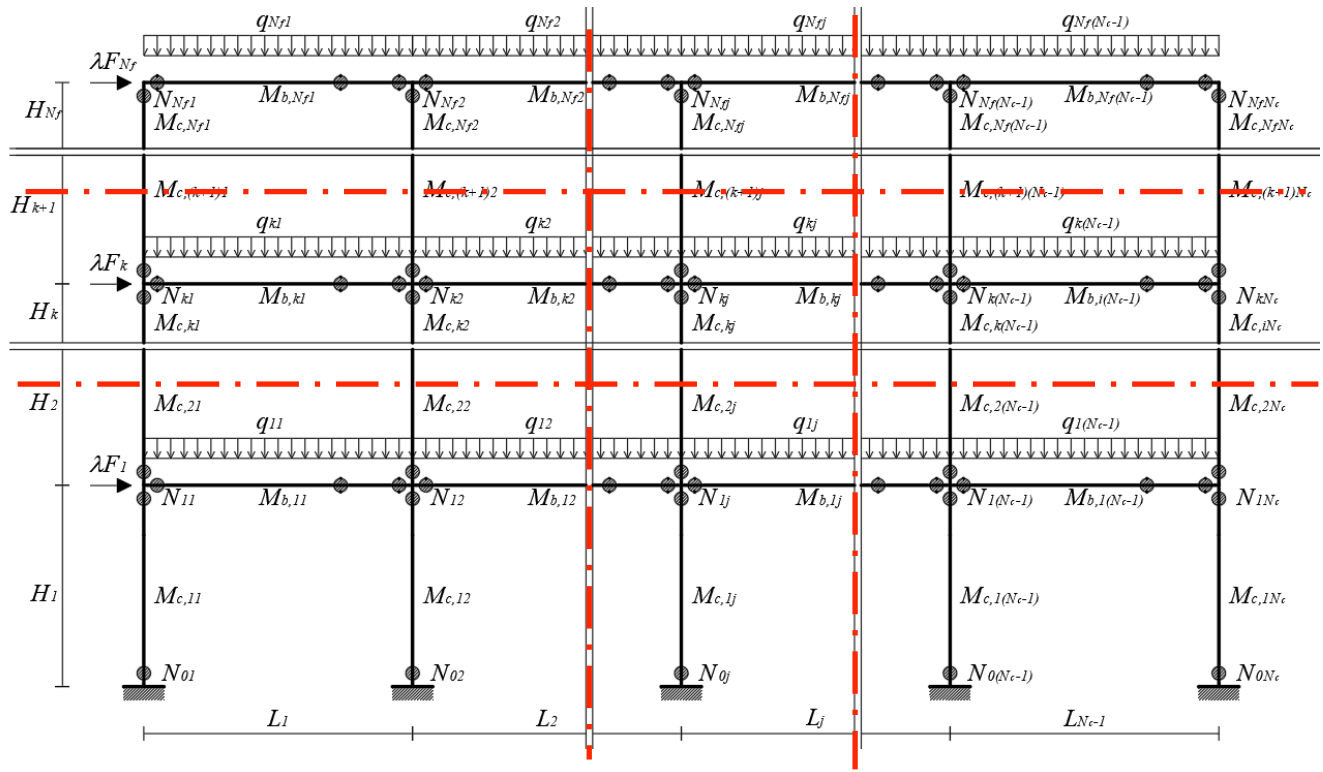

Figure 1. Layout of a generic planar frame

The optimization procedure adopted in the present paper makes use of genetic algorithms and populations of chromosomes are considered. Each individual of the population is coded as a string of integer numbers, where each number (called gene) represents how many times a given elementary mechanism enters in the combination. Therefore, if there are in total $N$ elementary mechanisms (obtained as the sum of floor, beam and node mechanisms), a generic chromosome $C_i$ of the population ($i = 1, ..., P$) represents a generic weighted combination of those mechanisms and can be coded in the following string:

$$C_i \quad (c_1, c_2, c_3, ..., c_k, ..., c_N) \tag{3}$$

with $c_k \in [0, c_{max}]$, being $c_{max}$ the maximum number of times the *k*-th mechanism is involved in the combination ($c_k = 0$ means that the mechanism is not involved at all). Given the string, it is possible to calculate the load factor $\lambda_0$ of the corresponding combination (chromosome). The overall number of possible different chromosomes is thus $P_{max} = (c_{max} + 1)^N$, a quantity which rapidly increases with *N* even for small values of $c_{max.}$. Task of the genetic algorithm is that of exploring the space of all the possible chromosomes, in search of the combination of mechanisms which minimizes $\lambda_0$ or, that is the same, which maximizes an appropriate "fitness function", here defined as:

$$f(\lambda_0) = K - \lambda_0 \tag{4}$$

where *K* is an arbitrary constant, chosen great enough to have $f > 0$ for every possible value of $\lambda_o$. The details of the application of the genetic algorithms together with the description of the original code developed in NetLogo multiplatform environment can be found in [16].

*2.2 Second order effects in the limit analysis of frame structures*

For each collapse complete mechanism, obtained combining the elementary ones, the second order effects of the work done by the vertical loads can be computed. This work, denoted as $dW_{extV}^{II}$ takes the following expression:

$$dW_{extV}^{II} = \sum_{i=1}^{N_f} \sum_{j=1}^{N_b} q_{ij} L_j h_i \varphi d\varphi \tag{5}$$

where $N_b = N_c - 1$ is the number of beams in each floor and $q_{ij}$ are the distributed vertical loads. Furthermore, $\varphi$ is a reference rotation with respect to which the whole failure mechanism is identified, $L_j$ is the length of the beams, and $h_i$ is a representative height (in the following addressed as *height of the collapse mechanism)*, related to the number of floor mechanisms, given by

$$h_i = h_{i-1} + H_i \alpha_i \tag{6}$$

where $H_i$ is the height of the columns at floor $i$ and $\alpha_i$ is equal to 1 for rotating columns, 0 otherwise (see Figure 2).

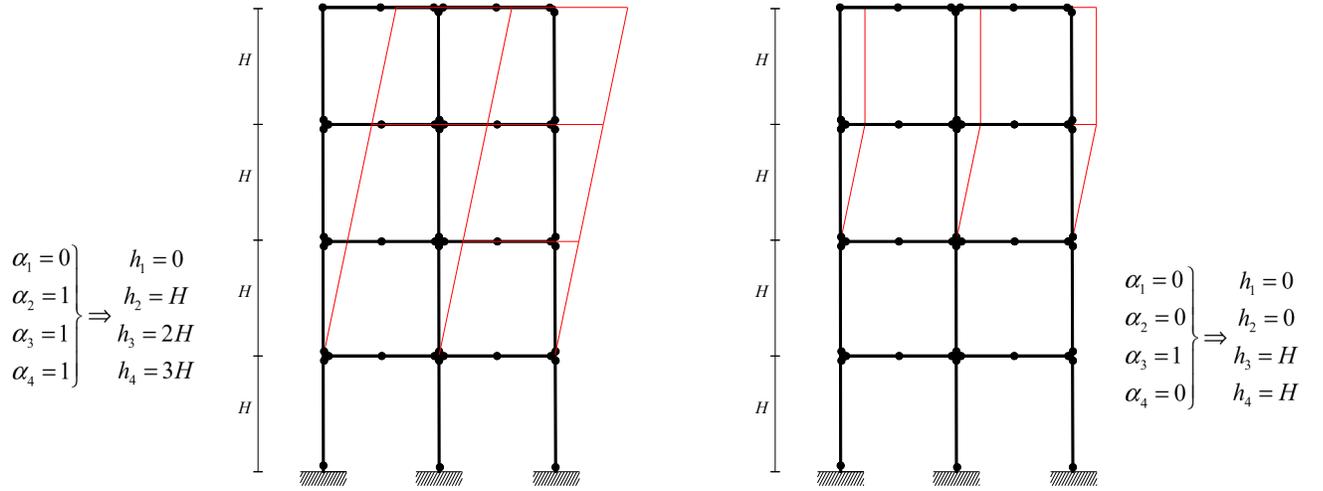

Figure 2. Evaluation of the height of collapse mechanisms, two representative cases

The second order work done by the vertical loads modifies, with respect to the linear analysis, the value of the collapse load multiplier as follows:

$$\lambda_c = \frac{W_{int} - W_{extV}^I - dW_{extV}^{II}}{W_{extH}} = \lambda_o - \gamma\delta \tag{7}$$

where

- $\lambda_o$ can be computed according to Eq. (2);

- $\gamma\delta = \dfrac{dW_{extV}^{II}}{W_{extH}} = \dfrac{\sum\limits_{i=1}^{N_f}\sum\limits_{j=1}^{N_b} q_{ij} L_j h_i \varphi d\varphi}{\sum\limits_{i=1}^{N_f} F_i h_i}$ is the decrease in the collapse multiplier due to second order effects;

- $\delta = \sum_{i=1}^{N_f} h_i \varphi = h_{max} \varphi$ is the horizontal displacement of the top of the frame;

- $h_{max}$ is the maximum value of $h_i$, evaluated according to Eq. (6).

The relation between horizontal displacement $\delta$ of the top of the frame, purged of the elastic deformability of the frame, and the load multiplier allows the reconstruction of a linear post-elastic descending branch accounting for the second order effects, as shown in Figure 3.

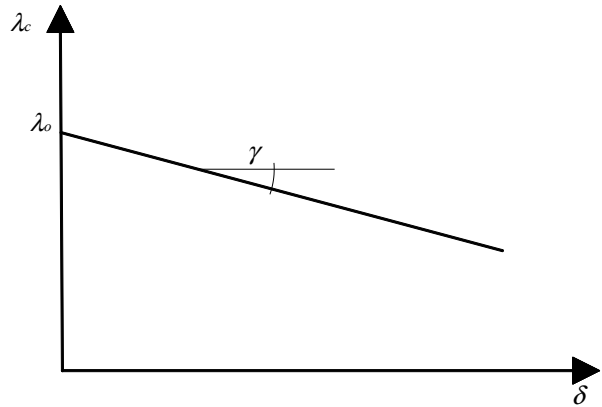

Figure 3. Load multiplier vs horizontal top displacement (purged of the elastic deformability) relation for a generic frame

## 3 Seismic performance of frames based on limit analysis

In this section a novel methodology, which exploits the outcome of limit analysis for a proper seismic vulnerability assessment of frame structures, is proposed.

Modern criteria (commonly accepted by many codes [2,3]) for the seismic assessment of frame structures, especially with reference to existing buildings, verify that the displacement capacity of an elastic-plastic single degree of freedom, which is considered 'equivalent' to the whole building, is compatible with the seismic demand associated to a design spectrum.

The equivalent single degree of freedom system is usually defined post-elaborating the outcome of nonlinear static analyses. The choice of adopting such a strategy is related with the need of defining a convenient ductility of a structure, which is usually associated to different criteria depending on the structural typology. Usually, at least for frame structures, the achievement of a base shear reduction as well as of the ultimate rotations in plastic hinges are accounted for [2,3]. Performing nonlinear static analyses requires a significant computational effort, since incremental steps (in terms of force or displacements) have to be applied, and for each of them a linear system has to be solved many times until the convergence is achieved.

Limit analysis was generally discarded as analysis strategy for the definition of an equivalent single degree of freedom for two main reasons:

i. it is based on the assumption of rigid perfectly plastic constitutive behaviour and the second order effects are usually neglected, thus making difficult the assessment of a softening branch after the peak load;

ii. the failure mechanism is associated to an increment of the field of displacement which cannot be quantified.

Assuming that the reconstruction of the post-peak softening branch is possible and that the displacement capacity of the frame can be reliably computed, limit analysis might represent a useful tool for the seismic assessment of structures. In addition, this strategy might present significant computational advantages with respect to nonlinear static analyses. In this section an attempt to interpret the outcome of the limit analysis for the definition of an approximated capacity curve is performed. Then, a proposal to define the displacement capacity of the structure on this equivalent single degree of freedom is made.

*3.1 Definition of an approximated capacity curve of the structure*

The approximated capacity curve here adopted is a bilinear one, referred to a monitored point conveniently chosen at the top floor of the frame. Such bilinear curve is defined through the computation of three main parameters: elastic stiffness, resistance, post-elastic slope. In the following these three parameters are defined as follows:

- the elastic stiffness $k_e$ is obtained by solving once a multi-degree of freedom linear system $\mathbf{u} = \mathbf{K}^{-1} \cdot \mathbf{f}$, (being $\mathbf{K}$ the stiffness matrix, $\mathbf{f}$ the load vector and $\mathbf{u}$ the elastic displacements), and then dividing the overall base shear $\sum_{j=1}^{N_f} F_j$ by the displacement of the monitored point $u_e$ (chosen among the terms of the vector $\mathbf{u}$), that is $k_e = \sum_{j=1}^{N_f} F_j / u_e$;

- the resistance $V_{by}$ is obtained multiplying the overall base shear and the collapse load multiplier $\lambda_o$ computed in subsection 2.1 $V_{by} = \lambda_o \sum_{j=1}^{N_f} F_j$;

- the post-peak slope $k_s$ is evaluated considering the second order effects, as described in subsection 2.2, and is equal to $k_s = -\gamma \sum_{j=1}^{N_f} F_j$.

It is worth to note that the yielding displacement can be computed as $u_y = \lambda_o \sum_{j=1}^{N_f} F_j / k_e = \lambda_o u_e$. The construction of this approximated bilinear curve is depicted in Figure 4. The capacity curve is then truncated considering as ultimate displacement $u_u$ the lowest among those associated to the achieving of limit states corresponding to several commonly accepted criteria. In the following subsections two widely accepted criteria to identify the ultimate conditions of frame structures are considered, namely the global base shear reduction (section 3.2) and the achieving of ultimate chord rotations in beams (section 3.3). The achieving of the ultimate conditions according to the two mentioned criteria is associated to two displacements of the approximated pushover curve, $u_s$ for the base shear reduction and $u_c$ for the achieving of the chord rotation on a beam. Finally, the ultimate displacement $u_u$ can be computed as

$$u_u = \min\{u_s, u_c\} \qquad (8)$$

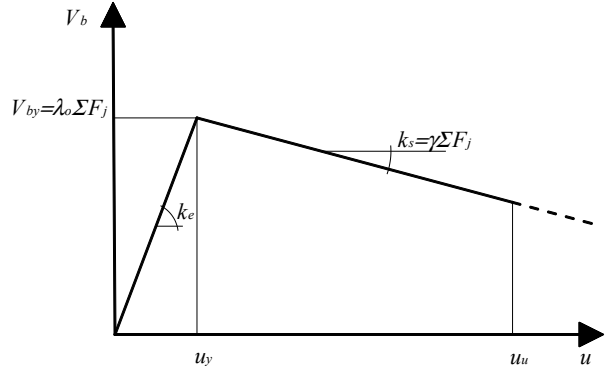

Figure 4. Approximated bilinear capacity curve

To validate the re-construction of the capacity curve through the proposed strategy the simple frame shown in Figure 5 is considered. The frame is a two bay and two storey one, with storey height equal to 3 m and equally spaced 4 m bays. The beams have a cross section corresponding to a HE300A profile at the first storey and to a HE240A one at the second storey, and are subjected to a permanent distributed vertical load equal to 50 kN/m. The columns are associated to cross sections corresponding to the profiles HE320A and HE240A at the first and second levels, respectively. The values of the plastic moments for each structural element are reported in the figure. Axial and shear deformabilities are neglected. The Young's modulus adopted for the steel is equal to 210000 MPa. The considered plastic moments are depicted in Figure 5. After assigning the vertical loads two horizontal force distributions are considered, namely a mass proportional force distribution and an inverse triangular one. The relevant horizontal forces are summarized in Table 1.

Table 1. Horizontal force distributions (in kN).

| | Force distribution | |
|---|---|---|
| | Mass proportional | Inverse triangular |
| $F_1$ | 400 | 266,67 |
| $F_2$ | 400 | 533,33 |

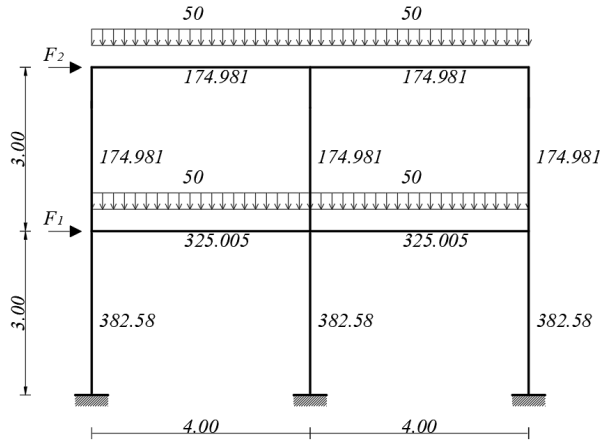

Figure 5. Benchmark planar frame (length in m, forces in kN)

The comparison between the proposed simplified strategy and the nonlinear static analyses here conducted by means of the well-known software SAP2000 [26], which employs a lumped plasticity approach, are reported in Figures 6 and 7. The FE numerical analyses account for large displacements, consistently with the limit analysis strategy here proposed. The right top corner of the second floor has been has been chosen as control point for the monitored displacement.

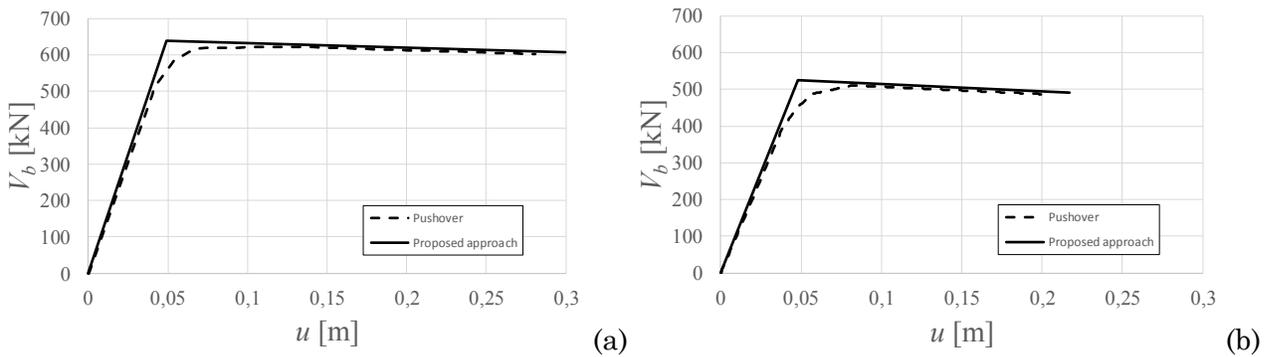

Figure 6. Comparison in terms of capacity curves between the nonlinear static approach and the proposed strategy for the frame depicted in Figure 5 considering two cases: (a) mass proportional and (b) inverse triangular load distributions.

In terms of capacity curves for both the considered load distributions, Figure 6 shows that the proposed strategy is in good agreement with the nonlinear static analyses both in terms of initial stiffness and with regard to the slope of the post-peak branch. The proposed strategy slightly overestimates the peak load; however, this is consistent with the fact that the nonlinear static analyses is able to grasp the second order effects also in the elastic phase. As expected, the triangular load distributions exhibit a lower resistance than the mass proportional one.

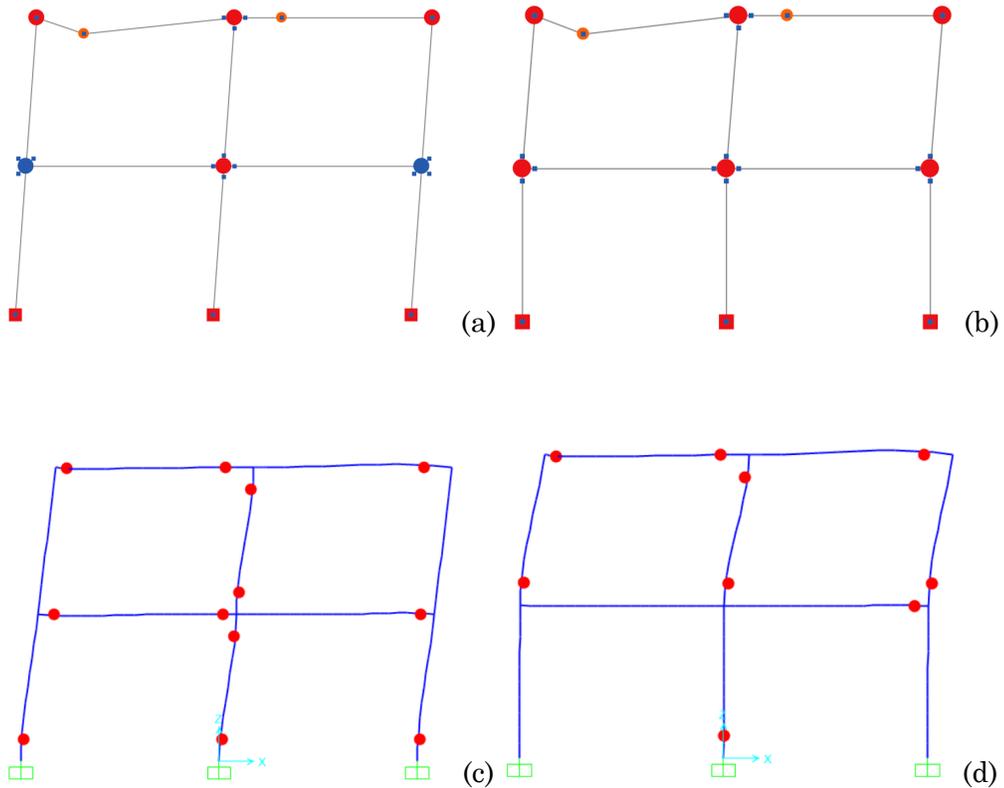

Figure 7. Comparison in terms of collapse mechanisms between the nonlinear static approach (c,d) and the proposed strategy (a,b) for the frame depicted in Figure 5 considering two cases: (a,c) mass proportional and (b,d) inverse triangular load distributions.

In terms of the collapse mechanisms, shown in Figure 7, it can be observed that the mass proportional distributions implies a global failure mode, whereas the inverse triangular load distribution provides a failure mode involving only the second storey. It is worth to mention that, although in the FE model the number of opened plastic hinges is greater than that strictly necessary for the failure mechanism, those involved in the failure mode correspond to those observed with the proposed approach.

*3.2 Displacement capacity associated to the base shear reduction*

Once a limit $\alpha_s$ in the base shear reduction with respect to the maximum achieved resistance is adopted (e.g. $\alpha_s = 0.15$ according to the Italian standards [25]), the corresponding limit of the monitored displacement $u_s$ can be computed considering the contribution associated to the elastic field and the one related to the post-peak branch whose loss of resistance is due to the second order effects, as in the following formula:

$$u_s = u_y + \frac{\alpha_s \lambda_o}{\gamma} = \lambda_o u_e + \frac{\alpha_s \lambda_o}{\gamma} = \lambda_o \left( u_e + \frac{\alpha_s}{\gamma} \right) \tag{9}$$

*3.3 Displacement capacity associated to the ultimate chord rotation of beams*

The method of the combination of elementary mechanisms does not allow a straightforward evaluation of the rotation in the plastic sections, nevertheless, it is possible to directly relate the plastic rotations in a given section with the displacement of the monitored point.

The following rule is here adopted to uniquely identify the plastic rotation at a given section. According to the numbering of nodes previously introduced, a generic node is identified with indexes $i$ (storey) and $j$ (column alignment); in a regular frame a maximum of four members can converge in such a node, that is the two beams identified with indexes $i,j$-1 (left) and $i,j$ (right), and the two columns identified with indexes $i$-1,$j$ (bottom) and $i,j$ (top). Each of the four sides of the node can be affected by a potential plastic rotation. For sake of conciseness, for each node the four plastic rotations can be identified as $\varphi_{ij,l}$, $\varphi_{ij,r}$, $\varphi_{ij,b}$, $\varphi_{ij,t}$, see also Figure 1.

As an example, if a column rotates around its clamped base, and considering the first column alignment, the plastic rotation associated to a given value of the monitored displacement $u$ can be computed as follows:

$$\varphi_{01,t} = \frac{u - u_y}{h_{max}} \tag{10}$$

In this case the computation is immediate, but a general procedure able to compute the plastic rotation in a generic hinge associated to a generic failure mechanism and magnitude of displacement is more complicated. For a generic hinge of a generic node, the magnitude of such plastic rotations depends on the contribution to the failure mechanism of five elementary mechanisms, that is: the relevant node mechanism, the two beam mechanisms associated to the two horizontal members converging on the node, and the two floor mechanisms associated to the bottom and top floors with respect to the node. For this reason, for each node a convenient sub-chromosome can be defined as follows:

$$C_{ij} = \left( c_{ij,n}, c_{ij,l}, c_{ij,r}, c_{ij,b}, c_{ij,t} \right) \tag{11}$$

where $c_{ij,n}, c_{ij,l}, c_{ij,r}, c_{ij,b}, c_{ij,t}$ represent the terms of the chromosome of Eq. (3) associated to the node mechanism, to the left beam, to the right beam, to the bottom and top floors, respectively. It is worth noting that $\varphi_{ij,r}$ can be whether the rotation at the node or at the along span hinge depending on the value of the acting vertical loads on the beam on the right of the considered node [19]. Once a reference rotation $\varphi_r(u)$ is related to the displacement of the capacity curve with the following formula:

$$\varphi_r(u) = \begin{cases} 0 & u \leq u_y \\ \dfrac{u - u_y}{h_{max}} & u > u_y \end{cases} \tag{12}$$

the plastic rotations of the four sections of the generic node are given by:

$$\varphi_{ij,l}(u) = \varphi_r(u)\left(c_{ij,n} + \frac{x_{i(j-1)}}{L_{j-1} - x_{i(j-1)}} c_{ij,l}\right)$$

$$\varphi_{ij,r}(u) = \begin{cases} \varphi_r(u)(c_{ij,n} - c_{ij,r}) & \text{if} \quad q_{ij} < \dfrac{4M_{b,ij}}{L_j^2} \\ \varphi_r(u)\left(\dfrac{L_j}{L_j - x_{ij}} c_{ij,r}\right) & \text{if} \quad q_{ij} \geq \dfrac{4M_{b,ij}}{L_j^2} \end{cases} \quad (13)$$

$$\varphi_{ij,b}(u) = \varphi_r(u)(c_{ij,n} - c_{ij,b})$$

$$\varphi_{ij,t}(u) = \varphi_r(u)(c_{ij,n} - c_{ij,t})$$

where $x_{ij}$ represents the distance between the along axis plastic hinge of a generic beam from its left end.

The displacement capacity of the structure can be associated to the achievement of the ultimate rotation in one of the plastic sections. To this purpose, the recent codes [2,3,25] introduce the concept of chord rotation, which has to be clarified in the case of limit analysis strategy. The chord rotation $\theta$ of a generic member is defined as the ratio between the difference of the transversal displacements at the plastic section and at the section where the moment vanishes and their distance $L_v$ (shear span), that is the ratio moment/shear at the plastic section. It is worth to note that the chord rotation accounts for both elastic and inelastic strain contributions; on the other hand, the relative rotations that can be evaluated according the proposed approach, Eqs. (13), consider the plastic rotation only. The limits usually adopted to identify the achieving of limit states in frame structures proposed in the codes [2,25] are usually a conventional yielding chord rotation $\theta_y$ (associated to the Limit State of Damage Limitation) and a ultimate chord rotation $\theta_u$ (associated to the Limit State of Near Collapse). For steel structures, according to [25], the yielding chord rotation $\theta_y$ is computed as follows:

$$\theta_y = \frac{M_p L_v}{2EI} \quad (14)$$

and with the ultimate rotation $\theta_u$ is evaluated according to the following formula:

$$\theta_u = 8\theta_y \tag{15}$$

In Eq. (14) $M_p$, $E$ and $I$ represent the plastic moment, the Young's modulus and the moment of inertia of a generic member, respectively.

Here, since the limit analysis approach cannot account for the elastic contribution of the rotations, the post-yielding contribution only will be considered. The ultimate condition is therefore identified when the plastic rotation in a generic hinge achieves a rotation equal to $7\theta_y$. According to this approach a correct evaluation of the shear span at the plastic sections is crucial. Here three possible scenarios, summarized in Table 2, can occur:

i. If the hinge occurs in a column $L_v$ is equal to half of the interstorey height for both the top and bottom hinges, $L_{vc,i} = H_i/2$;

ii. If the hinge occurs in a beam and the along length hinge does not open, that is $q_{ij} \leq 4M_{b,ij}/L_j^2$ [19], Figure 8a, the shear span of the right hinge $L_{vr,ij}$ can be seen as the distance between the right end of the beam and the abscissa where the bending moment vanishes and is $L_{vr,ij} = (2-\sqrt{2})\sqrt{M_{b,ij}/q_{ij}}$; then, the shear span $L_{vl,ij}$ of the left hinge is

$$L_{vl,ij} = L_j - L_{vr,ij}$$

iii. If the hinges occur in a beam and the along length hinge opens, that is $q_{ij} \leq 4M_{b,ij}/L_j^2$ [19], Figure 8b, the shear span of the right hinge can be computed as in the previous point, and that of the left hinge is then $L_{vl,ij} = L_j - x_{ij} - L_{vr,ij}$.

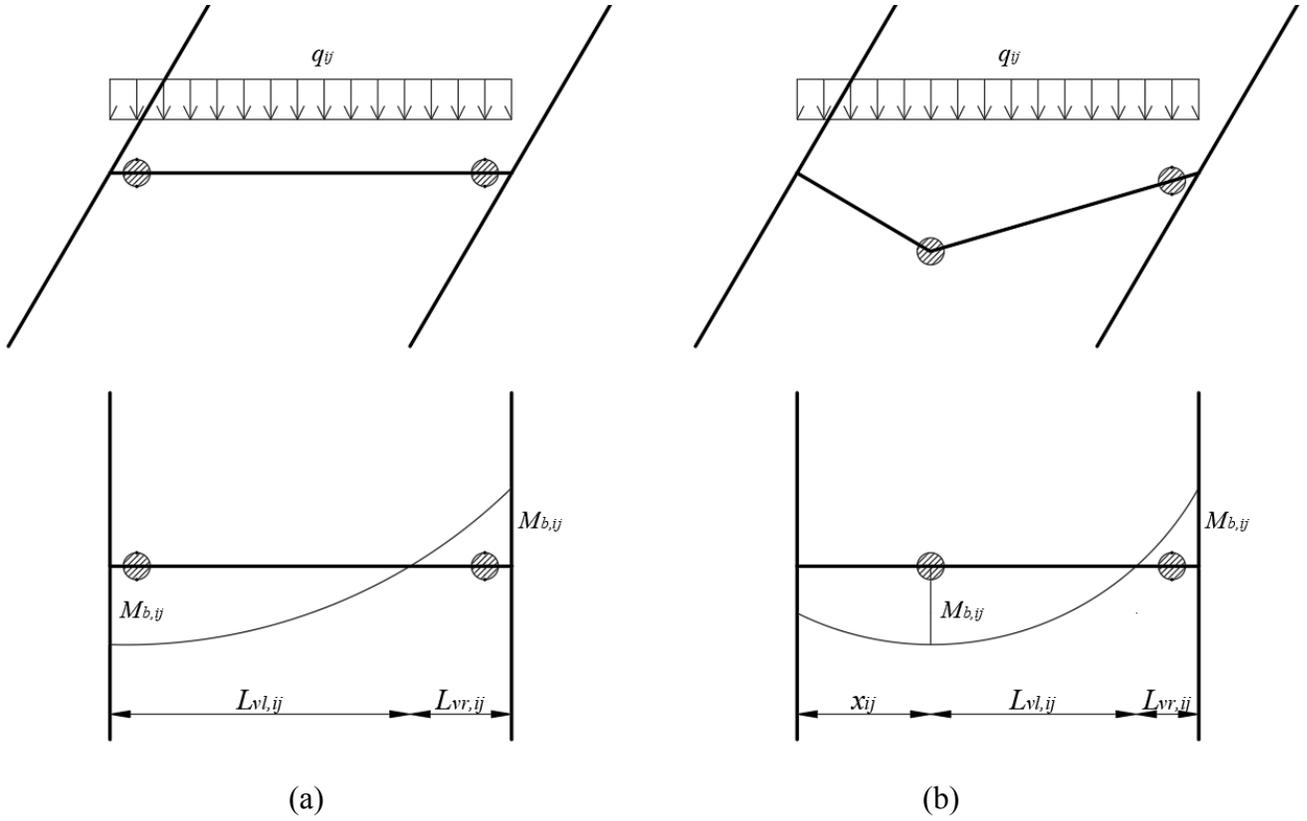

Figure 8. Shear span configurations in beams: (a) no intermediate hinge and (b) presence of intermediate hinge.

Table 2. Shear span associated to plastic hinges.

| Case | $L_{vl,ij}$ (left hinge) | $L_{vr,ij}$ (right hinge) |
|---|---|---|
| Column | $L_{vc,i} = H_i/2$ | $L_{vc,i} = H_i/2$ |
| Beam (no intermediate hinge) | $L_{vl,ij} = L_j - L_{vr,ij}$ | $L_{vr,ij} = \left(2-\sqrt{2}\right)\sqrt{\dfrac{M_{b,ij}}{q_{ij}}}$ |
| Beam (with intermediate hinge) | $L_{vl,ij} = L_j - x_{ij} - L_{vr,ij}$ | |

Once the ultimate rotation for a generic hinge is computed, the displacement of the capacity curve associated to the achievement of the ultimate rotation in a section $u_{ij,l}$ (or alternatively $u_{ij,r}$, $u_{ij,b}$ or $u_{ij,t}$) can be computed combining Eqs. (12) and (13) as follows:

$$u_{ij,l} = u_y + \dfrac{7\dfrac{M_{b,i(j-1)}L_{vr,i(j-1)}}{2EI_{b,i(j-1)}}}{c_{ij,n} + \dfrac{x_{i(j-1)}}{L_{j-1} - x_{i(j-1)}}c_{ij,l}} h_{max}$$

$$u_{ij,r} = \begin{cases} u_y + \dfrac{7\dfrac{M_{b,ij}L_{vl,ij}}{2EI_{b,ij}}}{c_{ij,n} - c_{ij,r}} h_{max} & \text{if} \quad q_{ij} < \dfrac{4M_{b,ij}}{L_j^2} \\[2em] u_y + \dfrac{7\dfrac{M_{b,ij}L_{vl,ij}}{2EI_{b,ij}}}{\dfrac{L_j}{L_j - x_{ij}}c_{ij,r}} h_{max} & \text{if} \quad q_{ij} \geq \dfrac{4M_{b,ij}}{L_j^2} \end{cases}$$

$$u_{ij,b} = u_y + \dfrac{7\dfrac{M_{c,ij}L_{vc,i}}{2EI_{c,ij}}}{c_{ij,b} - c_{ij,n}} h_{max}$$

$$u_{ij,t} = u_y + \dfrac{7\dfrac{M_{c,(i+1)j}L_{vc,i}}{2EI_{c,(i+1)j}}}{c_{ij,t} - c_{ij,n}} h_{max} \qquad (16)$$

Finally, the ultimate displacement $u_c$ of the capacity curve, associated to the achieving of the ultimate rotation in the first plastic hinge, is simply the minimum value of the displacements computed according to Eqs. (16).

*3.4 Validation of the proposed methodology*

In this section reference will be made to the benchmark frame depicted in Figure 5, whose refined and approximated capacity curves for two load combinations are reported in Figure 6. The latter curves have been truncated according to the criteria described in sections 3.2 and 3.3. In Table 3 the main data of the capacity curves are reported and, whenever possible, the error between the two approaches was included as well.

Table 3. Comparison of the main parameters of the capacity curves between nonlinear static analysis and the proposed approach.

|  | Mass proportional load distribution | | | Inverse triangular load distribution | | |
| --- | --- | --- | --- | --- | --- | --- |
|  | Nonlinear static analysis | Proposed approach | Error [%] | Nonlinear static analysis | Proposed approach | Error [%] |
| Peak load [kN] | 623,195 | 639,6848 | 2,58 | 511,074 | 524,1074 | 2,49 |
| $u_u$ [m] | 0,2809 | 0,2995 | 6,21 | 0,2014 | 0,2170 | 7,19 |
| $\gamma$ | - | 0,167 | - | - | 0,250 | - |
| $\lambda_o$ | - | 0,7996 | - | - | 0,6551 | - |
| $u_y$ | - | 0,0492 | - | - | 0,048 | - |

The two approaches show agreement in terms of the plastic hinges which lead to the achievement of the collapse condition for both load distributions. In particular, the collapse condition for the mass proportional load distribution is associated to the achieving of the ultimate chord rotation reached at the base of the central column. For the inverse triangular load distribution, the ultimate chord rotation occurs at the hinge located at the top of the central column at the second storey of the frame. It is worth to mention that, according to the proposed approach, the ultimate condition is simultaneously achieved in several hinges, which is consistent with the hypothesis of rigid members. However, additional checks have been made to verify that the ultimate displacements associated to the achievement of the ultimate chord rotations in plastic hinges located elsewhere, have magnitudes computed according to the proposed approach and to the performed nonlinear static analyses whose difference is comparable with those reported in Table 3 (maximum difference equal to about 7%); all the considered results are reported in Table 4. It is worth to note, that the hinge located at the left side of the left beam of the second storey, does not achieve the ultimate chord rotation, even for very large values of the displacement of the monitored displacement; this is consistent with the fact that such a hinge is located very close to the left end of the beam (about 25 cm), thus strongly limiting there the relative plastic rotation.

Table 4. Comparison in terms of achievement of the chord rotations between nonlinear static analysis and the proposed approach for two different load distributions with regard to the frame depicted in Figure 5.

| Hinge location | Mass proportional load distribution | | | Inverse triangular load distribution | | |
|---|---|---|---|---|---|---|
| | $u_u$ [m] (nonlinear static analysis) | $u_u$ [m] (proposed approach) | Error [%] | $u_u$ [m] (nonlinear static analysis) | $u_u$ [m] (proposed approach) | Error [%] |
| Left column, 1st storey (base) | 0,3142 | 0,2995 | 4,91 | - | - | - |
| Central column, 1st storey (base) | 0,2809 | 0,2995 | 6,21 | - | - | - |
| Right column, 1st storey (base) | 0,2942 | 0,2995 | 1,77 | - | - | - |
| Central column, 1st storey (top) | 0,3075 | 0,2995 | 2,67 | - | - | - |
| Left beam, 1st storey (left) | 0,5076 | 0,4953 | 2,48 | - | - | - |
| Right beam, 1st storey (right) | 0,3009 | 0,3150 | 4,48 | - | - | - |
| Left column, 2nd storey (base) | - | - | - | 0,2194 | 0,2170 | 1,11 |
| Central column, 2nd storey (base) | 0,3896 | 0,3873 | 0,59 | 0,2084 | 0,2170 | 3,96 |
| Right column, 2nd storey (base) | - | - | - | 0,2164 | 0,2170 | 0,28 |
| Central column, 2nd storey (top) | 0,3676 | 0,3873 | 5,09 | 0,2014 | 0,2170 | 7,19 |
| Left beam, 2nd storey (left) | 0,6742 | 0,7038 | 4,21 | 0,3523 | 0,3752 | 6,10 |
| Left beam, 2nd storey (right) | $\to \infty$ | $\to \infty$ | - | $\to \infty$ | $\to \infty$ | - |
| Right beam, 2nd storey (right) | 0,4009 | 0,3837 | 4,29 | 0,2174 | 0,2170 | 0,18 |

A further step that can be performed is a proper vulnerability assessment of the structure based on the N2 method [1], which employs an elastic perfectly plastic SDOF system equivalent to the structure to be inferred according to the capacity curves and to its modal properties. In the following the procedure proposed in the Italian Code [25] is employed. In particular, the adopted spectrum is computed considering the data associated to the city of Catania (Italy), soil type B, modal damping equal to 5%. The considered spectrum which is associated to a PGA equal to 0,283 g, is reported in the following Figure 9.

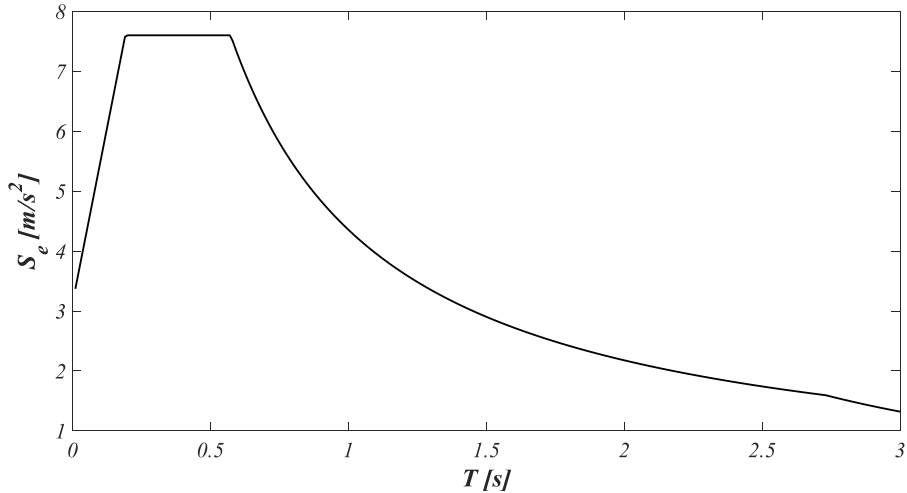

Figure 9. Spectrum adopted to assess the seismic vulnerability.

According to the procedure reported in [25], the properties of the equivalent SDOF and the demand and capacity associated to the Near Collapse (NC) Limit State are compared in Table 5. The superscript asterisk in the properties in the table refers to the equivalent SDOF.

Table 5. Comparison in terms of seismic vulnerability nonlinear static analysis and the proposed approach for two different load distributions with regard to the frame depicted in Figure 5.

|  | Mass proportional load distribution | | Inverse triangular load distribution | |
| --- | --- | --- | --- | --- |
|  | Nonlinear static analysis | Proposed approach | Nonlinear static analysis | Proposed approach |
| $\Gamma$ [-] | 1,1927 | 1,1932 | 1,1927 | 1,1932 |
| $F^*_u$ [kN] | 513,25 | 522,13 | 415,64 | 425,15 |
| $m^*$ [kN*s$^2$/m] | 53,43 | 53,74 | 53,43 | 53,74 |
| $d^*_y$ [m] | 0,0418 | 0,0402 | 0,0404 | 0,0389 |
| $d^*_u$ [m] | 0,2354 | 0,2510 | 0,1687 | 0,1819 |
| Seismic demand [m] | 0,3243 | 0,3082 | 0,3870 | 0,3670 |
| Safety factor | 0,7258 | 0,8143 | 0,4359 | 0,4955 |

The results reported in Table 5 show that the proposed procedure is able to provide safety factors close to those obtained with a standard approach employing nonlinear static analyses; in particular, the difference between the two approaches in terms of safety factor is around 12%. The proposed procedure is based on the assumption that all the plastic hinges simultaneously open when the collapse multiplier is achieved; on the other hand, the nonlinear static analysis is able to grasp the progressive damage occurring in the frame. The more the openings of such hinges occur close to each other, the more the two procedures provide similar results; it is worth to mention that the proposed procedure is not meant to be a thorough method to assess the seismic vulnerability of structures but rather as a preliminary assessment. In fact, due to its computational efficiency it could be employed for large scale vulnerability assessments or for design purposes comparing different frame configurations.

## 4  Conclusions

Limit analysis, although widely used for the evaluation of the resistance and the collapse mode of structures, is commonly not included among the tools that can be reliably employed for the seismic assessment of structures, since the evaluation of the ductility capacity cannot be easily computed. In this paper this belief is questioned and a procedure for the evaluation of a simplified bilinear capacity curve is proposed. To this purpose, the method of the combination of elementary mechanisms, already employed by the authors in combination with an evolutionary algorithm, is here extended to account for the second order effects, to be associated to a post-peak slope in the simplified capacity curve. An original strategy to evaluate the ultimate displacement associated to the achieving of the collapse condition according to several criteria commonly adopted for frame structures (base shear coefficient reduction and ultimate chord rotation) is then illustrated. The proposed strategy is validated and compared with a standard pushover approach considering different load scenarios. Although simplified, the proposed approach presents clear computational advantages with respect to a nonlinear static one, still guaranteeing an adequate level of precision. For the mentioned reasons the proposed approach could be reliably employed for preliminary seismic assessments or vulnerability evaluation of frame structures at the urban scale.